# Spontaneous Symmetry Breaking during Formation of ZnO Nanocrystals

Yan Zhou,[1,2,3] Junyan Zhang,[2,*] Jiangong Li,[1,*] and Bin Zhang[2]

[1]*Institute of Materials Science and Engineering, Lanzhou University, Lanzhou 730000, China*
[2]*State Key Laboratory of Solid Lubrication, Lanzhou Institute of Chemical Physics, Chinese Academy of Sciences, Lanzhou 730000, China*
[3]*Tianhua Institute of Chemical Machinery & Automation, Lanzhou 730060, China*

**ABSTRACT** Systemic symmetry is introduced into our work to investigate the symmetries of different ZnO nanocrystals. Result shows that systemic symmetries obey the law of spontaneous symmetry breaking during the formation of ZnO nanocrystals. A unitary formation mechanism of ZnO nanocrystals is proposed accordingly. Our results provide new insight into both crystallization and spontaneous symmetry breaking at meso-scale.

## 1. Introduction

Till now, despite various efforts to apply traditional theories of crystal growth to nanocrystals, some nanostructures still baffled researchers by their formation processes. In the case of ZnO nanocrystals, diverse growth mechanisms are proposed based on conjectures; however, not all of them are generally acceptable or ubiquitous, and some are even irreconcilable with each other. Since symmetry and symmetry breaking considerations dominate modern fundamental physics, we try to introduce it into investigating the formation mechanism for ZnO nanocrystals. Results show that the crystallization at meso-scale is completely different from those processes discovered by conventional crystallographic theories, and spontaneous symmetry breaking possesses the flexible priority over other principles to be satisfied during the formation of ZnO nanocrystals. This different proposition disposes spontaneous symmetry breaking to another unavoidable notion in crystallography at meso-scale.

Since ZnO nanocrystals exhibited the most splendid and abundant configurations, some of them, including tetrapod [1], rod-based tetrapod [1] and aeroplane-like structures [2] [Fig. 1], were selected as our objects of study. The tetrapod is a multi-twinning crystal [Fig. 1(a)], and the growth model widely accepted is that the sphalerite nuclei forms first, then the tetrapod structure results from the formation of wurtzite crystals protruding from the four plus {111} (including (111), $(\bar{1}1\bar{1})$, $(\bar{1}\bar{1}1)$ and $(1\bar{1}\bar{1})$) surfaces as four legs [3]. For the rod-based tetrapod [Fig. 1(b)], a completely different model is proposed [Fig. S1] that one dimensional ZnO nanorod forms firstly, then the side epitaxial growth of the nanorod results in the three sheets from $\{\bar{1}2\bar{1}0\}$ surfaces, finally four slender nanowires extend out from the three tips of the nanosheets and the end of the nanorod to generate a rod-based tetrapod structure [1]. Yet this proposition cannot explain why the structure is not a single crystal but a 4-polycrystal with special orientations. The aeroplane-like structure [Fig. 1(c)] has not been explained till now [2]. Although they were all fabricated by vapor deposition and resembled each other in shape, these ZnO nanocrystals have nothing in common about their growth mechanisms. However, the situation became different while spontaneous symmetry breaking was introduced into this research, for result showed that all these ZnO nanocrystals had intrinsic connection and they were intermediates of a common process.

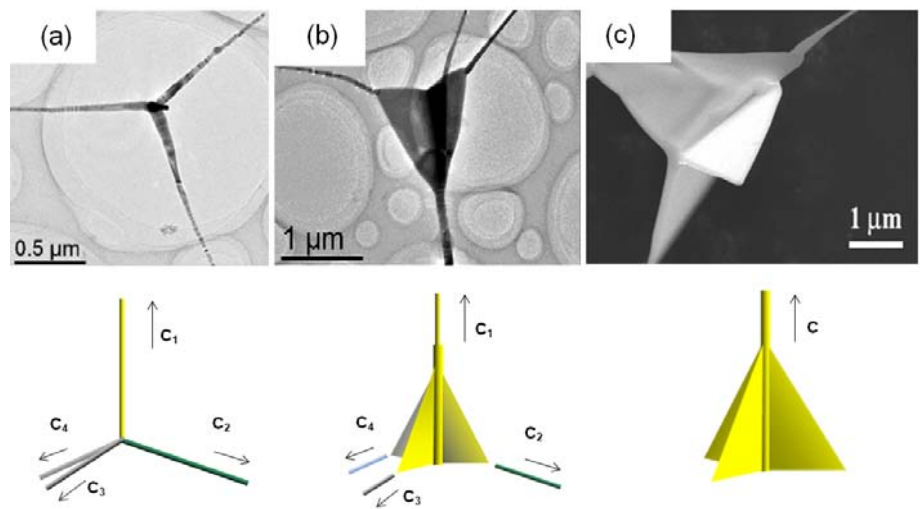

**Fig. 1.** Three morphologies of ZnO nanostructures and their corresponding models. (a) A tetrapod structure [1]. (b) A rod-based tetrapod structure [1]. (c) An aeroplane-like structure [2].

## 2. Symmetry breaking analysis

Before stating the common process mentioned above, two terminologies should be briefly introduced. One is the systemic symmetry. Although it often refers to the symmetries of unit cells and the crystal lattice when symmetries are involved in crystallography, the symmetries mainly discussed here are not the usual symmetries but rather symmetries of the figure and structure of the whole nano particle. For the ZnO nanocrystals shown in Fig. 1, the existence of symmetrical elements, such as symmetrical/anti-symmetrical centers, symmetrical axes and symmetrical planes, could not be denied. These fuzzy symmetries are named as systemic symmetries, and they are the subjects of the spontaneous symmetry breaking which is introduced as follows. Spontaneous symmetry breaking (SSB) occurs naturally in many situations. When some parameter reaches to a critical value, the symmetric equilibrium configurations become unstable and several equivalents with lower energy appear, then system transforms to one of the possible asymmetric stable equilibrium configurations [4]. SSB of systemic symmetries was found in our work to be the critical role which dictated the formation of ZnO nanocrystals. Thus, a hypothetical mechanism about SSB of systemic symmetries is proposed here (see Supplementary Information **b**). Fig.2 illustrates this mechanism, and the real structures [1, 2, 5] are quoted as evidence.

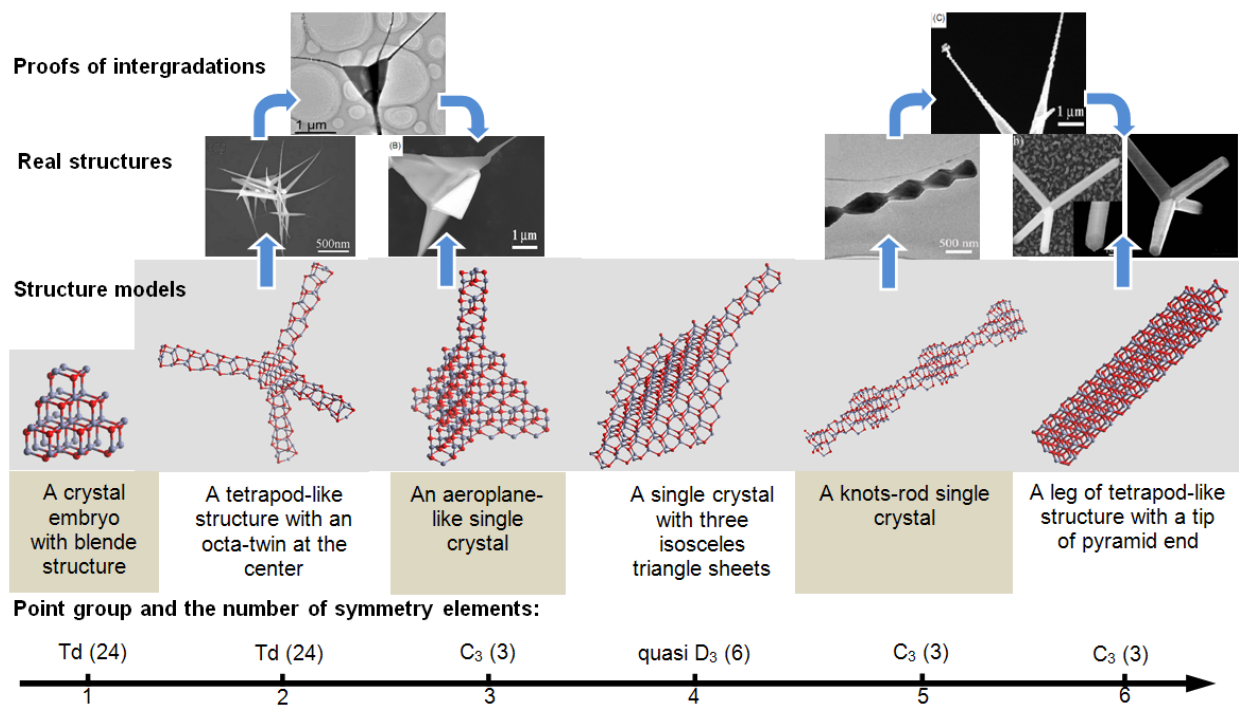

**Fig. 2.** Point groups degrade with symmetry breaking evolution during the formation of ZnO crystals.

From the perspective of SSB evolution of systemic symmetries, the formation of ZnO crystal could be divided into 6 stages, and each stage is labeled with a crystal lattice model [Fig. 2]. These models here reflect only the crystal orientations except the scale, size or defects of the real structures.

At stage 1, a cluster of ZnO with fcc structure is formed as an embryo [3], and its symmetry belongs to Td point group. Although ZnO with hcp structure is more stable under the given condition, the crystal embryo exists with fcc structure for that clusters tend to be with higher symmetry [6].

At stage 2, the hcp ZnO tends to form tetrapod legs by introducing stacking faults on the four plus {111} surfaces of the embryo [3]. Indeed, the transformation to hcp phase is subsequently driven by the higher free energy of fcc structures, so the fcc centre is hard to be found in any tetrapod structure. It should be noticed that the twins but not single crystals are usually obtained during the transformation, for Dai *et al.* proved the existence of octa-twin nucleus with hcp structure at the center of a ZnO tetrapod, and the intersections between any two legs are twinning planes [7]. Inheriting the symmetry of embryo formed at stage 1, the point group of a regular tetrapod still belongs to Td at this stage (see Supplementary information **a**3).

At stage 3, being driven by the high energy of twin planes (see Supplementary information **a**4), the orientation of one of the four legs dominates the whole structure of a tetrapod, and the other three legs recrystallize and change their orientations via shear transformation of lattice. At the end of this stage, a single crystal with three wings, which is called aeroplane-like structure [2], forms at the expense of the recessive twin crystals. The symmetry of a certain structure is breaking and degrading from Td to $C_3$ point group here, and this stage is detailed in later section [Fig. 3].

At stage 4, since the planes with Miller indices of {0001} are the planes with the fastest growth velocity [1], the top plane and the bottom plane in an aeroplane structure will grow preferentially and the final profile of the crystal will be limited by the facets with lower energy. At the end of this stage, a 3-wing structure, which is just like two aeroplanes connecting to each other with shared common bottom, could be supposed to be formed.

At stage 5, the corner between any two wings provides a concave site at which the vapor phase deposits preferentially. The image of a knotted-rod confirms this stage [Fig. 2].

At stage 6, the spaces between any adjacent knots form another kind of concave sites which tend to adsorb the vapor phase preferentially and be filled up to form the facets with lower energy lastly (see Supplementary information **a**5). The image between stage 5 and 6 shows the intergradations from the knotted-rod to the smooth rod. Tri-prism or hexa-prism with flat end or pyramid end forms at the end of this stage.

Although this hypothesis could not be proved completely, its first half explains the formation of rod-based tetrapod and aeroplane-like structure [Fig. 3]. The grain boundary shown in Fig. 3 is the most critical evidence which proves the formation mechanism of SSB while denies Gong's hypothesis.

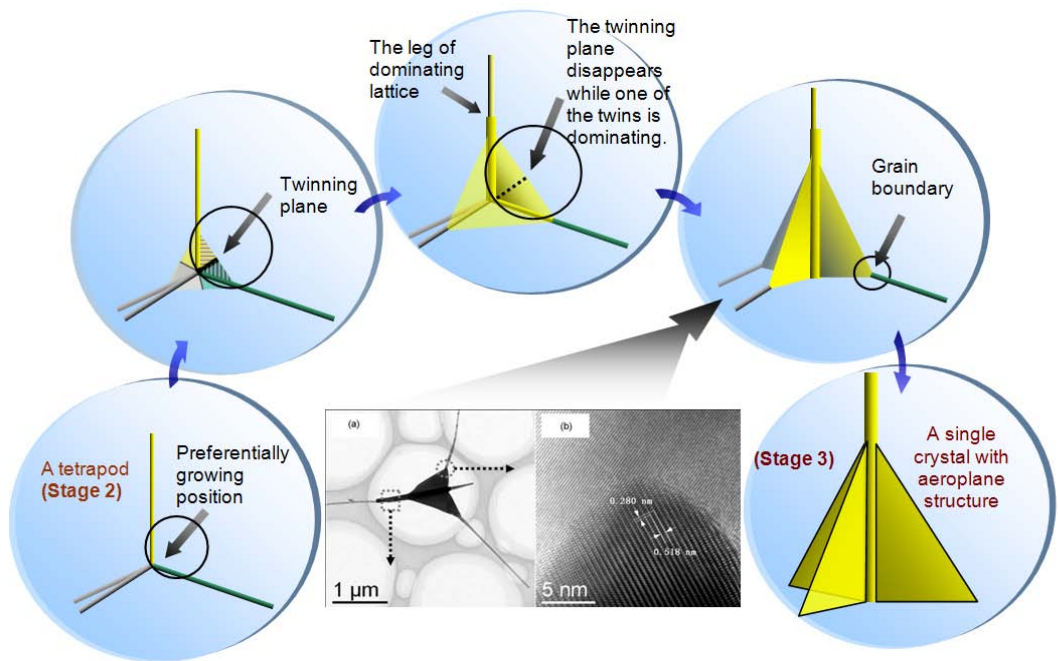

**Fig. 3.** A scheme of the transformation from Stage 2 to Stage 3 with real structure [1] as evidence.

### 3. Correlation between symmetry and potential

To discover the more profound significance of SSB during the process shown in Fig. 3, the correlation between symmetry and energy is investigated here. According to the concept of SSB [4], the energy of twinning planes is taken as the critical parameter to establish a calculation model [Fig. 4]. The volume of the twins and the energy of twinning planes can be calculated based on the calculation model:

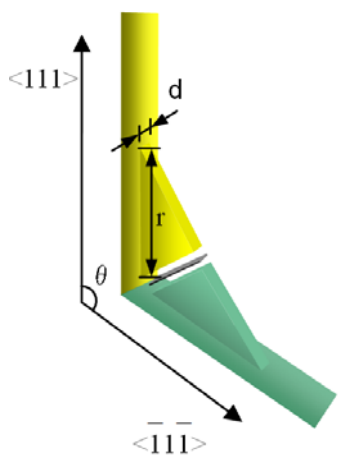

The parameters here are defined as follows:

$V$ ~ the volume of sheets

$U$ ~ the energy of twinned planes

$\sigma_t$ ~ the interfacial energy of twinned planes on per unit area

$d$ ~ the thickness of a triangular sheet, which is a constant depending on the thickness of the tetrapod's legs

$r$ ~ the side length of a triangular sheet

$\theta$ ~ the angle of 109.5° between any two adjacent legs in a regular tetrapod structure

$A$ ~ the area of the twinned planes

**Fig. 4.** A scheme with parameters illustrates one corner of a tetrapod structure.

$$V = \frac{1}{2} r^2 \cdot d \cdot \sin\theta$$

$$U = A\sigma_t = rd\sigma_t \cdot \cos\frac{\theta}{2} = \left(0.84\sqrt{d} \cdot \sigma_t\right) \cdot \sqrt{V}$$

For a certain ZnO nanoparticle, another parameter $\vec{r_i}$, which is defined as the position vector from the geometrical center to any unit cell, is involved in to denote the symmetry in a given structure. Phase diagrams of the topological character of symmetry can be established qualitatively besides the graph of V-U [Fig. 5]. Key points including A, B and C are labeled in U-V graph on the top of Fig. 5, and their corresponding states are detailed as follows. Firstly, point A reflects a state at which the structure is a regular tetrapod and the value $\sum_{Particle} \vec{r_i^n}$ (n=2N+1) remains 0 or near to 0, which suggests at least an approximate symmetry center in the structure at this stage [Fig. 5(a)]. Secondly, when the state develops from point A to point B, the area of twinned planes is increasing with the increased volume of the triangular sheets. Then, V and U come to the critical point of B, which demonstrates the maximum of U and the instability of the whole structure. The lattice of one of the four legs tends to be dominant in the whole structure when the value of U reaches its threshold of $U_c$. Fig. 5b shows that the topology of phase diagram bifurcates at point B, since the four directions are equal to each other. Lastly, the complete recrystallization results in a single crystal, and U drops to 0 for the disappearance of all twining planes. The state of a perfect single crystal lies at one of the four points in the phase diagram [Fig. 5(c)].

During the above process, when system develops to the state at key point B [Fig. 5], a fourfold degenerate states appears, for any of the four orientations of the tetrapod legs could be the dominating orientation, and that could lead to any of the four equivalent results. Here the existence of a fourfold degenerate states results in symmetry breaking. When one leg is dominating and the twinning planes begin to disappear, four possible equilibrium configurations with $C_3$ point group appears. From the perspective of statistics, the symmetry of system still belongs to Td point group and the symmetry of Lagrange's variable of system is never breaking. Yet for a certain structure, all of these four degenerate states could not be occupied simultaneously. Thus, when the real structure transforms to one of the possible configurations, symmetry is breaking from Td to $C_3$ point group, and a rod-based tetrapod forms. Regardless of a quantum particle or a nano particle, the intrinsic reason for SSB is that a certain system can only occupy one of the multifold states at a certain time.

The correlation of energy and symmetry can be demonstrated directly by a 2D and a 3D phase diagram [Fig. 5(d, e)], which show the transformation from a tetrapod, then a rod-based tetrapod, and finally to an aeroplane-like structure.

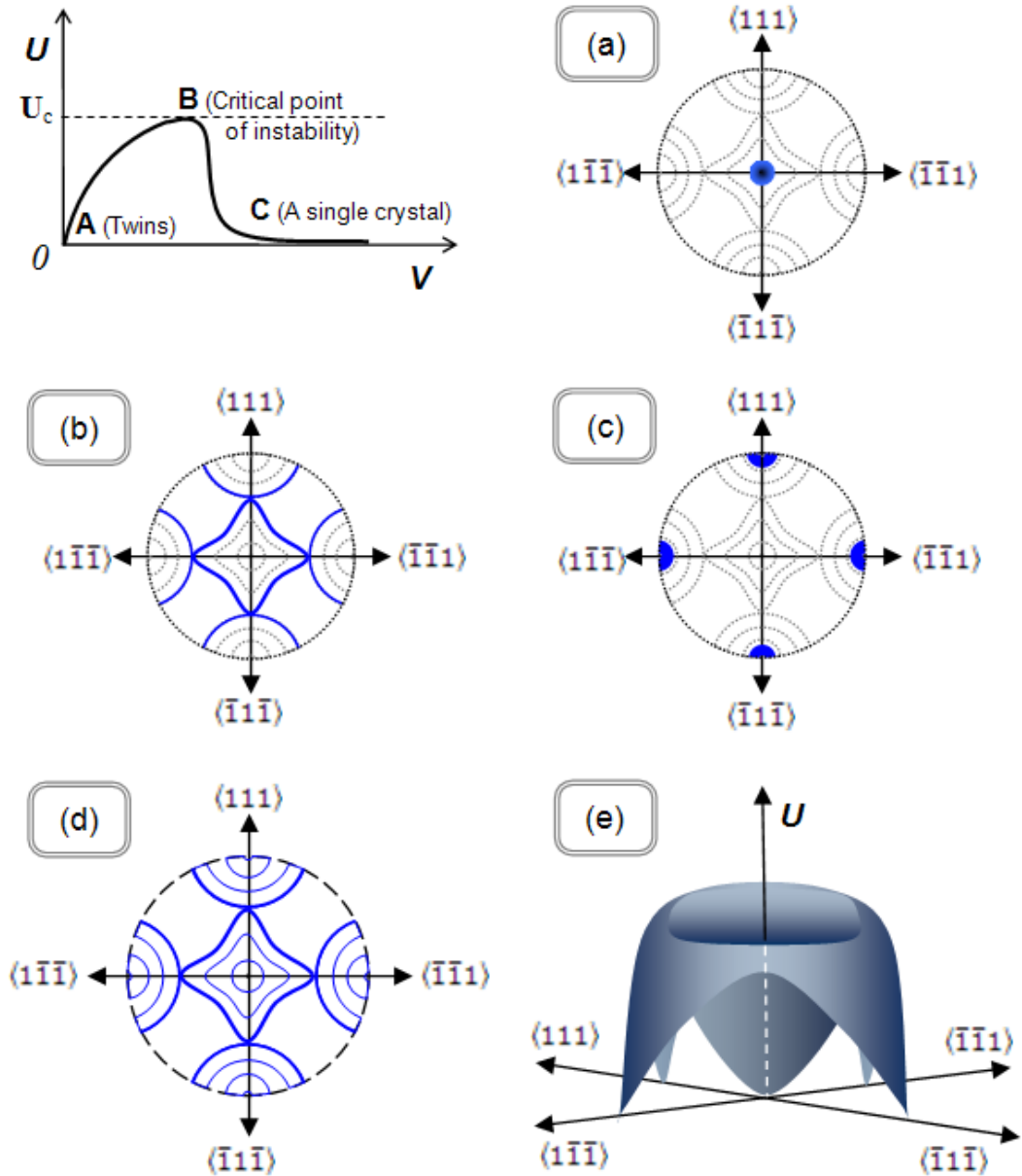

**Fig. 5.** The relationship between V and U. (a), (b) and (c) Phase diagrams (which are corresponding to point A, B and C in the U-V graph above, respectively) show the qualitative topological character of symmetry at each labeled state. (d) & (e) 2D and 3D phase diagrams demonstrate the topology of symmetry during the transformation from stage 2 to stage 3.

4. **Results and discussions**

This section summarizes the results of our investigation. During the formation of ZnO nanocrystals, SSB provides the key clue to establish a unitary formation mechanism for the structures including tetrapod, rod-based tetrapod and aeroplane-like structures. Some puzzling phenomena, such as the strange structure of rod-based tetrapod and the formation of aeroplane-like structure, could be explained directly by SSB. Furthermore, the grain boundary between the triangular sheet and its epitaxial wire, which is the greatest mystery during the formation of rod-based tetrapod, is explained automatically by our mechanism.

Fractal analysis is also involved here. As shown at the bottom right corner of Fig. 2, the rod with $C_3$ symmetry is just one leg of another tetrapod structure, which is much larger than the initial tetrapod from which the rod stems. The above phenomenon suggests that symmetry breaking plays its role hierarchically with fractal as a supplement.

### 5. Conclusions

Finally, this paper includes that SSB dictates the crystallization at meso-scale. Since SSB has been applied successfully in many fields such as quantum and subatomic physics, it seems more significant that SSB is helpful to probe crystallography at meso-scale. The formation of ZnO nanocrystal demonstrates irrefutably that SSB provides new insight into nano-structures, while fuzzy systemic symmetry provides a new clew to study nano-crystals. Our work also shows that the SSB dictates the world hierarchically and extensively, which is proved to be far more exceeding expectation again.

**Acknowledgements**

The authors greatly acknowledge the "Hundreds Talent Program" of the Chinese Academy of Sciences and NSFC (Grant No. 50283008, 50721062) for financial support. The cunning experiment and objective records made by Gong, *et al*. are also appreciated.

[*]Corresponding Author: Tel.: 86-931-4968295; Fax: 86-931-4968295.

E-mail address: junyanzh@yahoo.com; lijg@lzu.edu.cn

# Supplementary Information a

## Spontaneous Symmetry Breaking during Formation of ZnO nanocrystals

Yan Zhou,[1,2,3] Junyan Zhang,[2,*] Jiangong Li,[1,*] and Bin Zhang[2]

[1]*Institute of Materials Science and Engineering, Lanzhou University, Lanzhou 730000, China*

[2]*State Key Laboratory of Solid Lubrication, Lanzhou Institute of Chemical Physics, Chinese Academy of Sciences, Lanzhou 730000, China*

[3]*Tianhua Institute of Chemical Machinery & Automation, Lanzhou 730060, China*

**CONTENTS**

1. **Symmetry of ZnO crystals**

Comparing the fcc structure cluster or the octa-twin nucleus with an hcp single crystal, it is found that neither the cluster nor the octa-twin is more stable than an hcp single crystal at the given condition, whereas both of them are more symmetrical. As it plays a critical role in determining the stability of clusters and the existence of octa-twins, symmetry becomes an unavoidable factor to probe the formation of these ZnO nano crystals. The nucleation and growth of ZnO crystal both obey the law of symmetry breaking, for the symmetry of the crystal's structure degrades from Td to $C_3$ point group when ZnO grows from an embryo to an aeroplane-like single crystal.

2. **Gong's mechanism for rod-based tetrapod structure of ZnO nanocrystals**

In order to draw a more rational conclusion, the relevant opinion of Gong's [1] is demonstrated here as comparisons [Fig. S1].

FIG. S1. The growth model of ZnO with rod-based tetrapod structure.

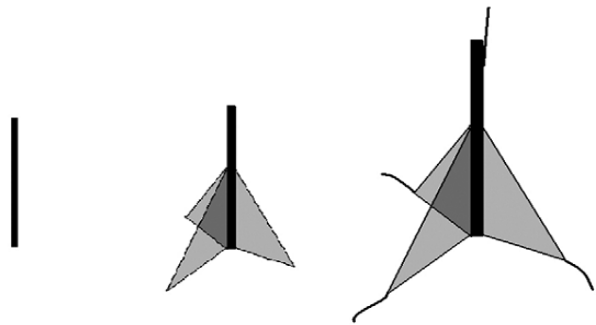

3. **Symmetry of octa-twins of ZnO**

If only the lattice of the octa-twins is concerned, the symmetry of the structure belongs to Oh point group. According to the result of Y. Dai *et al.*, the polarities of the twinned crystals are not mirror-symmetric but antisymmetric, for every twin is of the inversion type [Fig. S2(a)] [2]. Thus when the sign of the Miller indices is also concerned, the symmetry of the octa-twin will degrade from Oh point group to Td point group. It should be noted that the twinning planes could not be the coherent planes of $\{11\bar{2}2\}$, and the reason will be detailed in the next section.

FIG. S2. Multiple twins of tetrapod ZnO. (a) An octa-twin is composed of eight pyramidal inversion-twin crystals , and each of these pyramids is formed with three $\{11\bar{2}2\}$ and one (0001) facets. (b) 7-twins. (c) 6-twins. (d) 6-twins. (e) 4-twins.

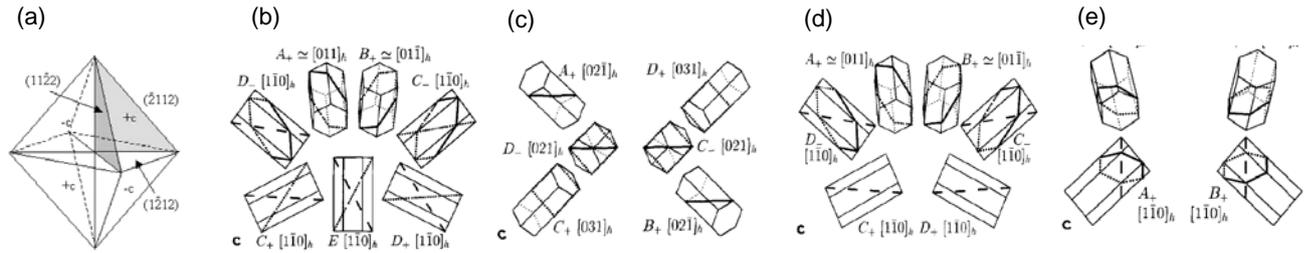

4. **An analysis of the crystal parameters of ZnO**

An analysis of crystal parameters is essential for explaining the formation of ZnO nanocrystals. The lattice constants of ZnO with hcp structure are a=0.3296 and c=0.52065 nm [1], therefore, the value of c/a is 1.580. Supposing that hcp ZnO is formed by stacking fault of fcc phase, the value of c/a should be 1.633, which is about 3% larger than the real value 1.580. This subtle difference suggests that distortion is involved in hcp structure during the stacking faults. Thus, for any given ion in hcp ZnO crystal, the four adjacent reverse ions should not lie at each corner of a regular tetrahedron. Without consideration of this distortion, any two hcp legs protruding from center will form twins by sharing coherent $\{11\bar{2}2\}$ plane as a common plane, as shown in Fig. S2(a). But this structure model has to be modified since the subtle distortion denies the possibility that the twinning-plane is coherent and without distortion. Thus, the energy of the twinning plane in real structure will be higher than that of coherent $\{11\bar{2}2\}$ twinning plane. Under given conditions, the higher energy may leads to two kinds of results, one is that the real twinning planes tend to disappear completely via recrystallization, and it is confirmed by the single crystal character of aeroplane-like structure; the other is that the real twinning planes tend to turn into new ones with lower energy, and the tendency will destroy the structure of a regular octa-twins. The fact is that these two kinds of transformations take place simultaneously. Nishio K. et al. reported a series of tetrapod ZnO structures with centers including 7-twins, 6-twins, and 4-twins [3]. In our opinion, these different twin particles of ZnO all resulted from the two kinds of transformation. Some of the adjacent twins recrystallize and becomes a new one while the twinning plane between them disappears, and the elements of an octa-twins continues to decline from 8 to 7, 6, 4 and even to single. Figure S2 (b-e) shows the structures of these multiple twins. In addition, the twin direction changes and the symmetry decline during this transformation. The transformation from octa to single is discussed in this letter, while those twins between 8 and single are the results of incomplete transformation, and they will be discussed further in Section 5.

Whatever the final structure is, the high energy of the twin boundary is the root cause of transformation of octa-twins, and the crystal structure of ZnO is the intrinsic reason for the high boundary energy.

## 5. The side facets of the tri-prism or hexa-prism

The image [4] [Fig. S3] contains some information about the side facets of the tri-prism or hexa-prism. Geometrical analysis reveals that these side facets are $\{11\bar{2}0\}$ planes. FIG. S3. SEM image of one leg in a tetrapod ZnO. The inset is from the top view of the leg [4].

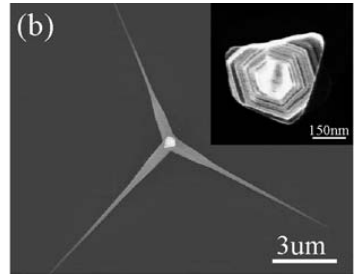

## 6. Energy analysis of incomplete symmetry breaking in the formation of ZnO crystals

Besides the transformation from an octa-twin to a single crystal, there are some other tendencies during the formation of ZnO twin crystals, including 7, 6, 4-twins [Fig. S2]. According to the calculation model shown in the letter, the energy of the twin planes in these twins could be illustrated in Figure S4.

FIG. S4. The relationship between the V and U.

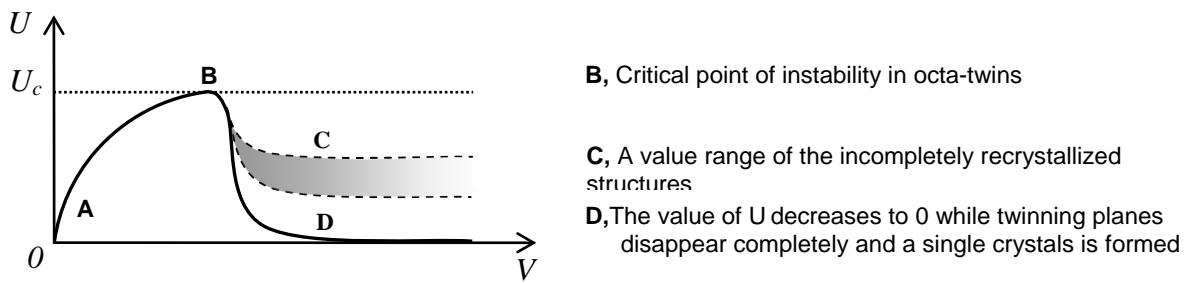

**B,** Critical point of instability in octa-twins

**C,** A value range of the incompletely recrystallized structures

**D,** The value of U decreases to 0 while twinning planes disappear completely and a single crystals is formed

## 7. Potential application to other tetrapod nanocrystals

Some other tetrapod nanocrystals have been fabricated recently, such as CdTe [5], CdS [6], ZnS [7], ZnSe [8], and CdSe [9]. Since Zn and Cd are both homologues and hcp structures in normal case, the mechanism based on symmetry breaking might make sense of other tetrapod-like structures.

# Supplementary Information **b**

Spontaneous Symmetry Breaking during Formation of ZnO nanocrystals

Yan Zhou,[1,2,3] Junyan Zhang,[2,*] Jiangong Li,[1,*] and Bin Zhang[2]

[1]*Institute of Materials Science and Engineering, Lanzhou University, Lanzhou 730000, China*

[2]*State Key Laboratory of Solid Lubrication, Lanzhou Institute of Chemical Physics, Chinese Academy of Sciences, Lanzhou 730000, China*

[3]*Tianhua Institute of Chemical Machinery & Automation, Lanzhou 730060, China*

**CONTENTS**

1. A PPt file containing Flashes shows our argument briefly and directly. See: http://cid-3c4b4badb40d55df.skydrive.live.com/redir.aspx?resid=3C4B4BADB40D55DF!106&authkey=XBU8Rd9Ax8Q%24